\documentclass[conference]{IEEEtran}
\IEEEoverridecommandlockouts
\usepackage{cite}
\usepackage{amsmath,amssymb,amsfonts}
\usepackage{algorithmic}
\usepackage{graphicx}
\usepackage{textcomp}
\usepackage{xcolor}
\usepackage{url}
\usepackage{comment}
\usepackage{graphicx}
\usepackage{fancyhdr}

\def\BibTeX{{\rm B\kern-.05em{\sc i\kern-.025em b}\kern-.08em
      T\kern-.1667em\lower.7ex\hbox{E}\kern-.125emX}}
\begin{document}

\title{Contract Wallet Using Emails}

\author{\IEEEauthorblockN{Sora Suegami}
\IEEEauthorblockA{\textit{ Dept. of Information and Communication Engineering} \\
\textit{The University of Tokyo}\\
Tokyo, Japan \\
suegamisora@g.ecc.u-tokyo.ac.jp}
\and
\IEEEauthorblockN{Kyohei Shibano}
\IEEEauthorblockA{\textit{Dept. of Technology Management for Innovation} \\
\textit{The University of Tokyo}\\
Tokyo, Japan \\
shibano@tmi.t.u-tokyo.ac.jp}
}
\IEEEpubid{\makebox[\columnwidth]{978-8-3503-1019-1/23/\$31.00~\copyright2023 IEEE \hfill} \hspace{\columnsep}\makebox[\columnwidth]{ }}
\maketitle
\IEEEpubidadjcol

\begin{abstract}
We proposed a new construction for contract wallets, smart contract applications that allow users to control their crypto assets.
Users can manipulate their crypto assets by simply sending emails with no need to manage keys.
These emails are verified using zero-knowledge proof (ZKP) along with their attached digital signatures that the sender domain server (SDS) generates according to DomainKeys Identified Mail.
Unless the SDS forges the emails, the crypto assets remain secure in the proposed system.
Moreover, the existing SDSs can be used as is by outsourcing additional work to a third party that is not necessarily trusted.
The system supports various functions to manipulate crypto assets.
We produced a tool for variable-regex mapping (VRM) that enables developers to build a new function without ZKP skills.
For example, using the tool, we built a demo application where users can exchange crypto assets via Uniswap only with emails.
\textbf{The published version of this paper is available at \url{https://doi.org/10.1109/ICBC56567.2023.10174932}.}
\end{abstract}


\section{Introduction}
In Ethereum blockchain, contract wallets hold users' crypto assets and provide more functional management of these assets than standard wallet services such as MetaMask.
For example, some contract wallets require multiple digital signatures to transfer crypto assets \cite{praitheeshan2020security}.
However, most of them require users to install new software or access new web pages.
In other words, users cannot manipulate their crypto assets with existing familiar tools, e.g., email.

To solve this problem, we proposed \textbf{Email Wallet}, a contract wallet that enables users to manage their crypto assets by simply sending emails.
Our system is based on a technique in \cite{zkmail} and \cite{ecdsa_nullifier:online} that verifies the sender domain of emails with zero-knowledge proof (ZKP).
The crypto assets stored in our system are secure unless the SDS forges the user's email, and the user needs to manage no private keys.
Additional work for our system can be outsourced to a third party called an aggregator, which is neither a user nor an SDS, and not necessarily trusted.

Our new technique, variable-regex mapping (VRM), allows a developer to build a new function to manipulate assets without ZKP skills.

\section{System overview}\label{section_system_overview}
Our system is one of the contract wallets operated by users, aggregators, and the Ethereum blockchain.
Users are identified by their email addresses. They deposit their crypto assets in a smart contract, called a wallet contract, by specifying their email addresses and transferring the assets to the contract address.
The deposited assets are manipulated according to functions defined as manipulation rules.
Each manipulation rule consists of an ID, a regular expression (regex), and another smart contract called a manipulation contract that implements the manipulation method.
The regex (e.g., ``Transfer \textbackslash d\{1,20\} wei Ether.'') is decomposed into fixed parts (e.g., ``Transfer'') and variable parts (e.g., ``\textbackslash d\{1,20\}''), and the manipulation contract has access to the latter values.
The user specifies the ID of the manipulation rule in the email title and writes the email body to match the corresponding regex.

\begin{figure}[hbtp]
      \centering
      \includegraphics[scale=0.24]{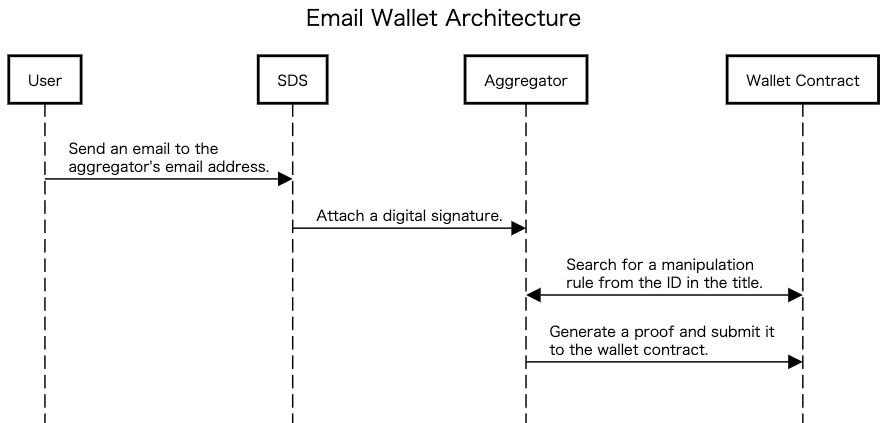}
      \caption{Email Wallet architecture}
      \label{fig:email_wallet_architecture}
\end{figure}

The user's email is processed by the SDS that supports DomainKeys Identified Mail (DKIM) and an aggregator (Figure \ref{fig:email_wallet_architecture}).
Herein, we assume that a public key of the SDS is published in the domain name system and is previously registered in the wallet contract.
We suppose the SDS follows pkcs1-v1\_5 defined as RFC 3447 \cite{RFC3447P15:online}: it computes a hash value of the email with SHA256 and generates an RSA digital signature for that value.
It is also assumed that the SDS does not change the public key or forge the user's email and the aggregator publishes its own email address.
\begin{enumerate}
      \item The user sends an email to the aggregator's email address.
      \item The SDS attaches a digital signature to the email according to the DKIM protocol.
      \item The aggregator uses the ID in the email title to search for a manipulation rule and checks that the email content satisfies the corresponding regex.
      \item The aggregator generates a proof with the received email. Then, it creates a transaction including the proof, sender's email address, RSA public key, ID, and values of the variable parts of the regex and submits it to the wallet contract.
\end{enumerate}
The wallet contract verifies the proof with the other submitted data and checks that the SDS's public key matches the one registered in the contract.
If the verification succeeds, the wallet contract invokes the manipulation contract corresponding to the ID to manipulate the crypto assets of that user.

\subsection{Implementation}
Our implementation consists of a ZKP circuit, smart contracts, and the tool for the VRM.
The circuit was built with the halo2 library developed by Privacy \& Scaling Explorations team \cite{halo2_library:online}.
The circuit takes the email header/body and the RSA digital signature as a witness (or private input) and the sender's email address, RSA public key, ID, and values of the variable parts of the regex as an instance (or public input).
It computes the hash value of the email using a SHA256 chip that supports variable-length input and a base64 conversion chip. Then, it verifies the signature using an RSA verification chip for the hash value and the public key.
It also uses a regex verification chip to verify that a string that combines the fixed parts with the values of the variable parts satisfies the regex.
The last chip is implemented according to the idea proposed in \cite{ecdsa_nullifier:online}. It transforms a regex into an equivalent finite automaton and verifies that the initial state is transitioned to an accepted state by a given string.
Notably, the maximum string length is fixed because the circuit size must be fixed.

Our smart contracts, i.e., the wallet contract and the manipulation contracts, are implemented with Solidity.
The wallet contract stores each user's email address and balance per currency unit as user information. It also maintains an ID and address of the manipulation contract for each manipulation rule.

The tool for the VRM is used to automatically generate a new regex verification chip.
Specifically, for each manipulation rule, a developer defines a regex and implements a manipulation contract with Solidity.
The tool outputs a bytecode of a verifier contract that verifies a proof for the circuit corresponding to the regex.
The developer only has to call that contract before manipulating the crypto assets, with no prior knowledge of ZKP.

\section{Demo application}
We developed a demo application according to the description in Section \ref{section_system_overview}.
There are two manipulation rules:
\begin{itemize}
      \item \textbf{Rule 1:} If a user with $X$ sends an email with the message ``Transfer $A$ $T$ to $Y$,'' the balance of $T$ for $X$ and $Y$ will be decreased and increased by $A$, respectively.
      \item \textbf{Rule 2:} If a user with $X$ sends an email with the message ``Swap $A$ $T_1$ to $T_2$ via Uniswap,'' the amount $A$ of $T_1$ in the balance of $X$ is exchanged for $T_2$ with Uniswap, and the balance of $T_2$ is increased.
\end{itemize}
Herein, $X$ and $Y$ denote email addresses; $A$ and $T$ represent a remittance amount and its currency unit, respectively.
Using Rule 1, the user can transfer crypto assets deposited in the wallet contract by specifying the balance/currency unit and the recipient's email address.
Furthermore, Rule 2 allows the user to exchange those assets for another currency using Uniswap.
The wallet contract address is used to send and receive assets to and from Uniswap.

Using the tool for the VRM, we obtained the regex verification chips and bytecodes of the verifier contracts necessary for Rules 1 and 2.

\subsection{Tests}
We tested our implementation using the following scenario.
There are two users, Alice and Bob.
Aditionally, there is a SDS and an aggregator.
First, Alice deposited 0.01 ETH in the wallet contract.
Alice then transferred 0.005 ETH to Bob's email address using Rule 1.
Bob converted 0.005 ETH to DAI with Uniswap using Rule 2.
We confirmed that Alice's final balance is 0.005 ETH and Bob holds only some DAI, which amounts to 0.005 ETH when exchanged on Uniswap.
We also tested invalid cases.
In cases where the attached digital signature was invalid or the email did not match any valid regexs, the submitted proof did not pass the verification and the users' balances were not modified.
These results show the tool for the VRM exports correct regex verification chips and the bytecodes for Rules 1 and 2.

\section{Conclusion}
In this study, we proposed a contract wallet that enables users to operate their crypto assets by simply sending emails.
The user only needs to trust existing SDSs without managing any private keys.
The user's crypto assets are manipulated according to various rules, and the VRM technique allows developers to build new rules without writing any ZKP circuits.
We developed a demo application and proved that the user can transfer crypto assets and exchange them for different currencies just by sending an email to an aggregator.

\section*{Acknowledgment}
This work has been supported by Endowed Chair for Blockchain Innovation and the Mohammed bin Salman Center for Future Science and Technology for Saudi-Japan Vision 2030 (MbSC2030) at The University of Tokyo.
We appreciate the generous technical advice and implementation help from Mr. Aayush Gupta.

\bibliographystyle{IEEEtran}
\bibliography{references.bib}

\begin{thebibliography}{1}
\providecommand{\url}[1]{#1}
\csname url@samestyle\endcsname
\providecommand{\newblock}{\relax}
\providecommand{\bibinfo}[2]{#2}
\providecommand{\BIBentrySTDinterwordspacing}{\spaceskip=0pt\relax}
\providecommand{\BIBentryALTinterwordstretchfactor}{4}
\providecommand{\BIBentryALTinterwordspacing}{\spaceskip=\fontdimen2\font plus
\BIBentryALTinterwordstretchfactor\fontdimen3\font minus
  \fontdimen4\font\relax}
\providecommand{\BIBforeignlanguage}[2]{{%
\expandafter\ifx\csname l@#1\endcsname\relax
\typeout{** WARNING: IEEEtran.bst: No hyphenation pattern has been}%
\typeout{** loaded for the language `#1'. Using the pattern for}%
\typeout{** the default language instead.}%
\else
\language=\csname l@#1\endcsname
\fi
#2}}
\providecommand{\BIBdecl}{\relax}
\BIBdecl

\bibitem{praitheeshan2020security}
P.~Praitheeshan, L.~Pan, and R.~Doss, ``Security evaluation of smart
  contract-based on-chain ethereum wallets,'' in \emph{International Conference
  on Network and System Security}.\hskip 1em plus 0.5em minus 0.4em\relax
  Springer, 2020, pp. 22--41.

\bibitem{zkmail}
\BIBentryALTinterwordspacing
S.~Suegami, ``Rsa verification circuit in halo2 and its applications - privacy
  and scaling explorations,'' 2022. [Online]. Available:
  \url{https://mirror.xyz/privacy-scaling-explorations.eth/mmkG4uB2PR_peGucULAa7zHag-jz1Y5biZH8W6K2LYM}
\BIBentrySTDinterwordspacing

\bibitem{ecdsa_nullifier:online}
A.~Gupta, ``An ecdsa nullifier scheme and a proof of identity application,''
  Master's thesis, Massachusetts Institute of Technology, September 2022.

\bibitem{RFC3447P15:online}
I.~N.~W. Group, ``Rfc 3447: Public-key cryptography standards (pkcs) \#1: Rsa
  cryptography specifications version 2.1,''
  \url{https://www.rfc-editor.org/rfc/rfc3447}, February 2003, (Accessed on
  01/01/2023).

\bibitem{halo2_library:online}
P.~.~S. Explorations, ``privacy-scaling-explorations/halo2,''
  \url{https://github.com/privacy-scaling-explorations/halo2}, (Accessed on
  01/01/2023).

\end{thebibliography}

\end{document}